# Metal – Insulator transition in $Fe_{1.01-x}Cu_xSe$

A J Williams[1], T M McQueen[1], V Ksenofontov[2], C Felser[2], and R J Cava[1]

[1] Department of Chemistry, Princeton University Princeton NJ 08544

[2] Institut für Anorganische Chemie und Analytische Chemie, Johannes Gutenberg-Universität, Staudinger Weg 9, D-55099 Mainz, Germany

**Abstract**

Iron Selenide, $Fe_{1.01}Se$, the layered parent compound of the recently discovered superconducting arsenide family, has previously been shown to be non magnetic and superconducting with a critical temperature of 8 K. Here we show that copper can be substituted at the iron site in $Fe_{1.01}Se$ up to a solubility limit of 20-30 %, after which a first order transition to the three-dimensional $CuFeSe_2$ structure type is observed. As little as 1.5 % percent copper is sufficient to suppress the superconductivity, and 4 % drives the system through a metal-insulator transition. A local magnetic moment is introduced, which maximizes near 12% doping, where a spin-glass transition near 15 K is observed.

**Introduction**

Superconductivity has recently been observed in iron arsenide-based compounds, with superconducting transition temperatures ($T_c$'s) as high as 55 K [1-5]. This discovery was unexpected, as most Fe-based compounds display magnetic ordering at low temperatures, and has ignited a fierce debate regarding the underlying physics. This iron arsenide family, and the previously reported $LaFePO_{1-x}F_x$ [6], share many structural features, each being made up of $Fe_2X_2$ (X = P, As) layers of edge-sharing $FeX_4$ tetrahedra, separated by either single metal ions [2,3,5] or metal-containing layers [1,4,6]. This same basic structural motif is also found in the comparatively simple tetragonal form of iron selenide (called β-FeSe, referred to as simply "FeSe" in this report), which has itself been shown to superconduct at 8 K [7]. While FeSe has $Fe_2Se_2$ layers that are isomorphic to the $Fe_2X_2$ layers described above, it lacks intermediate chemical substituents that can themselves affect the electronic and structural properties within the layers. It therefore represents an ideal compound in which to study the interplay of structure, magnetism and superconductivity within this superconducting materials family.

The extreme sensitivity of the superconductivity to stoichiometry in $Fe_{1+\delta}Se$ has previously been demonstrated [8], with the highest $T_c$'s being found when the material is closest to stoichiometric ($Fe_{1.01}Se$). $T_c$ is enhanced under high pressure conditions, reaching a maximum of ~37 K [9-11]. No long-lived magnetism is observed, but recent $^{77}Se$ NMR results [12] do provide evidence for short range spin fluctuations, potentially linked to the superconducting mechanism in this compound. FeSe has also been shown to undergo a low temperature structural modulation [13-14], analogous to the FeAs-based materials. Early reports suggested this to be a simple tetragonal-to-orthorhombic symmetry change, however recent electron microscopy results [15] suggest that this distortion may be more complex.

Despite the comparative simplicity of the FeSe system, its similarities to the high $T_c$ FeAs-based superconductors, and the fact that it has been known for 30 years [16-17], relatively few chemical substitution studies have been reported. The $FeSe_{1-x}Te_x$ solid solution has been the subject of intense recent study [18-19] owing to the enhancement of $T_c$ to around 14 K for x~0.5. A first order phase transition is, however, observed in this system at $0.15 \leq x \leq 0.3$ to another phase which, while retaining the PbO-type structure, has a much larger interlayer spacing. To our knowledge, only one previous study has been reported involving substitution at the iron site in FeSe [20]. In that work, $Fe_{1-x}Ni_xSe$ and $Fe_{1-x}Co_xSe$ were successfully prepared up to x = 0.2, and in each case this substitution was shown to suppress superconductivity. No transition to semiconducting behaviour was observed, however, and no magnetic measurements were reported.

We report here the synthesis and characterization of a novel solid solution series $Fe_{1.01-x}Cu_xSe$, in which copper substitutes directly for iron. X-ray powder diffraction evidences a linear dependence of the lattice parameters with Cu doping up to a solubility limit of x ~ 0.3, at which point a first order phase transition to the alternate $CuFeSe_2$ structure [21], which has copper between the $M_2X_2$ layers, is observed. Resistivity measurements show that superconductivity is rapidly suppressed by Cu-doping of only around 1.5 %, with further Cu doping producing a semiconducting state that persists, with an increasing resistivity, up to the solubility limit. Magnetisation measurements, combined with Mössbauer studies show that this semiconducting behaviour coincides with the appearance of a local moment in $Fe_{1.01-x}Cu_x$-Se.

**Experimental**

All samples were prepared from iron pieces (Johnson-Matthey, 99.98%), selenium shot (Alfa-Aesar, 99.999%) and copper foil (Alfa-Aesar, 99.9985%). Stoichiometric quantities of freshly polished iron and copper, and selenium shot were loaded into cleaned and dried silica tubes. A piece of cleaned carbon was placed at the opposite end of the tube (and prevented from coming into contact with the sample) and the tube sealed under vacuum. These tubes were then sealed inside a second evacuated silica ampoule, and placed in a furnace at 750 ºC. The temperature was slowly ramped up to 1075 ºC over the course of 4 days, and then held at that temperature for a further 24 hours. The temperature was then rapidly decreased to 420 ºC, held for an additional 48 hours, and then reduced to 330 ºC for a final annealing step of 2-5 days. Finally, the tubes were quenched into -13 ºC brine. All samples are stable for short periods of time in air, but were protected from oxidation by storage in an argon glovebox.

The polycrystalline samples obtained were studied by powder X-ray diffraction (XRD) using a Bruker D8 Focus employing Cu K$\alpha$ radiation and a graphite diffracted beam monochromator. Patterns for all samples were Rietveld analysed using the GSAS software package [22]. Temperature-dependent magnetization and electronic transport properties were measured in a Quantum Design physical property measurement system (PPMS). Resistivities were measured using the standard four-probe method, with silver paste cured at room temperature being used for the contacts. $^{57}$Fe Mössbauer spectra were recorded in a transmission geometry using a conventional constant-acceleration spectrometer and a helium bath cryostat. The Recoil Mössbauer Analysis Software was used to fit the experimental spectra.

**Results and Discussion**

Initial attempts to synthesise copper-doped iron selenide were based on the hypothesis that the copper would be intercalated between the FeSe layers, in an analogous fashion to what is found for CuFeSe$_2$ [21]. To this end, a series of samples with general stoichiometry Cu$_x$Fe$_{1.01}$Se was prepared (x ≤ 0.14) and studied by laboratory XRD. Close inspection of the diffraction patterns showed that the low Cu-doped samples appeared to be phase pure, whereas the higher Cu-doped samples were contaminated by increasing amounts of elemental iron. In order to more accurately quantify the levels of iron impurity in these samples, M(H) curves were measured at 150 K from 0 to 9 T (Fig. 1 inset) and compared against a sample of pure Fe powder. A systematic, near linear, trend of saturation magnetization with x was clearly observed (Fig. 1), and analysis indicated that the amount of iron impurity found in these samples was approximately equivalent to the amount of copper being added. The explanation is that copper is not intercalating in the van der Waals gap in the FeSe structure, but rather substituting at the iron position in the FeSe layers. A new set of samples with general stoichiometry Fe$_{1.01-x}$Cu$_x$Se was therefore prepared.

Figure 2a shows diffraction patterns for a full set of Fe$_{1.01-x}$Cu$_x$Se samples with 0 < x ≤ 0.5. All samples appear to crystallize in the same tetragonal cell, space group *P4/nmm*, as FeSe. Samples at low Cu-dopings contain small amounts of Fe$_7$Se$_8$ impurity, but no iron impurity is observed across the entire series. M(H) curves were again measured at 150 K from 0 to 9 T, as a further test for iron contamination, and the saturation magnetizations (as a proportion of the signal from elemental Fe) are also shown in Figure 1. Unlike Cu$_x$Fe$_{1.01}$Se, there is no systematic trend in the observed high field magnetization with x, and above x = 0.06, there is no evidence for any significant ferromagnetic impurities. Thus the magnetic and diffraction data indicate that the solid solution is of the type Fe$_{1.01-x}$Cu$_x$Se.

The diffraction data for all samples were Rietveld analysed, and the refined lattice parameters for $0 < x \leq 0.2$ are shown in Figure 3a. All samples were well fitted in space group *P4/nmm*, and no additional superstructure reflections were observed. As iron is replaced by copper in this solid solution, the in-plane lattice parameter increases systematically, whilst the inter-plane distance decreases. The overall effect of these lattice changes is to compress the $FeSe_4$ tetrahedra (Fig. 3b), opening up the Se-Fe-Se bond angle, and lengthening the FeSe bonds (The geometry of the $FeSe_4$ tetrahedra is influenced by the selenium z-coordinate, the only refinable atomic parameter within this space group. This coordinate did not change by more than $2\sigma$ across the whole series, and is therefore constant within the precision of our current measurements).

The sample at $x = 0.3$ is two phase. For $x = 0.4$ and $0.5$, the samples are single phase and have a diffraction pattern consistent with what is expected for $CuFeSe_2$ [21,23]. $CuFeSe_2$ adopts a larger unit cell (approximately $\sqrt{2}a \times 2c$ bigger than FeSe), with a similar structure to FeSe, but with 1/4 of all the metals transposed from the layers to sites between the layers (Fig.2b inset). While the superstructure reflections that evidence this supercell are not observable by laboratory X-ray diffraction, the subcell we observe for $Fe_{0.51}Cu_{0.5}Se$ (a = 3.9017(3), c = 5.520(1) Å) is in agreement with that of the published structure [21]. Furthermore, quantitative refinement of the data in this *P-42c* cell, with metal atoms positioned in between the layers, gives a substantially improved fit for the $x = 0.4$ and $0.5$ samples ($\chi^2$ for $Fe_{0.51}Cu_{0.5}Se$ drops from 3.093 to 1.486, $R_F^2$ from 21.72 % to 12.58 %). The diffraction pattern for $Cu_{0.3}Fe_{0.71}Se$ can be fitted (Fig. 2b) as a phase separated mixture of the FeSe and $CuFeSe_2$ structure types.

When taken together, the structural data indicate that copper can successfully be substituted for iron in FeSe up to a threshold of 20-30 % copper, after which a first order structural transition to the alternate $CuFeSe_2$ structure type is observed. Why this more three-dimensional structure should be favoured at high copper dopings, and moreover why the transition between the two closely related structures should be first order, is not clear at this time – further structural studies will be of interest to fully understand this behaviour.

The dependence of the sample resistivity with copper content, normalized to the value at 300 K for these polycrystalline materials, is shown in Figure 4. Undoped $Fe_{1.01}Se$ is superconducting, with a transition temperature of approximately 8 K. Even the addition of very small amounts of copper into the structure significantly lowers the $T_c$. The dc magnetization of $Fe_{0.995}Cu_{0.015}Se$ shows no evidence of bulk superconductivity (Fig. 4 inset). For x = 0.04, all semblance of a downturn in the resistivity curve has disappeared, and above this doping level, all samples are semiconducting, with the resistivity increasing systematically with x up to the solubility limit of this phase. It is clear that a small amount of electron doping into the $Fe_{1.01}Se$ structure has a very profound effect on the behaviour of this system. Only a couple of mole percent of copper are necessary to suppress any superconductivity in $Fe_{1.01}Se$ – consistent with previous work that demonstrated a similarly extreme dependence of superconductivity on stoichiometry in this material [8]. $Fe_{0.97}Cu_{0.04}Se$ represents the addition of only 0.12 valence electrons per mole, and yet this is already sufficient to tune the system from a metal to a semiconductor.

Measuring the normal state magnetic susceptibility of $Fe_{1.01-x}Cu_xSe$ is a more involved procedure, owing to the presence of small amounts of ferromagnetic iron impurity – which even at less than 0.1% concentrations, can have a profound effect on the low field magnetization [24]. In order to obtain a good estimate of the intrinsic magnetic susceptibility of these materials, the magnetization was measured between 1.8 and 300K separately under fields of 2T and 4T – above the field at which the magnetic Fe impurity saturates. Figure 5 shows a plot of the susceptibility, measured as M(4T)-M(2T)/ΔH, against temperature. The upper inset shows the behaviour of values of χ, extracted at 10K and 298K, with x. The magnetic susceptibility of $Fe_{1.01}Se$ decreases almost linearly from 300K down to ~100K, below which temperature it levels off, before exhibiting a small Curie tail - behaviour qualitatively consistent with previous $^{77}Se$ NMR measurements [12]. The Cu-doped samples have a somewhat reduced susceptibility at all temperatures relative to the undoped parent – a feature which we can attribute to a reduction of the density of states upon doping. With 2% Cu-doping, a small low temperature local moment is observed, which increases monotonically with the addition of more and more copper up to a doping level of 12%.

For $Fe_{0.89}Cu_{0.12}Se$, a downturn in susceptibility is observed at around 5 K, consistent with the formation of a spin glass. At higher copper contents, this downturn shifts to higher temperatures but also broadens substantially. The lower inset shows field cooled (FC) and zero-field cooled (ZFC) measurements of the susceptibility of $Fe_{0.89}Cu_{0.12}Se$ under a smaller field of 1 T (there are no ferromagnetic impurities present in this concentration range). The transition has shifted to a higher temperature (~ 15 K) in the ZFC measurement, but is not seen at all in the FC measurement. These observations are again consistent with the presence of glassy magnetic ordering, which is destroyed by field cooling in a 1 T field. Further characterisation of this spin glass phase may be of interest in future studies.

Mössbauer data were collected at 4.2K on samples $Fe_{0.99}Cu_{0.02}Se$, $Fe_{0.97}Cu_{0.04}Se$ and $Fe_{0.89}Cu_{0.12}Se$, and are displayed in Figure 6. For $Fe_{0.99}Cu_{0.02}Se$, the data can be described by a single quadrupole paramagnetic doublet, as for the parent $Fe_{1.01}Se$ [8]. However the spectra for the higher Cu-doped samples are more complex, and require the addition of a second sextet term in their fitting. This sextet, implying the presence of magnetic fluctuations in the system, is fitted with a site population of around 13% in $Fe_{0.97}Cu_{0.04}Se$ and 46% in $Fe_{0.89}Cu_{0.12}Se$.

While no evidence for magnetic ordering has previously been observed in $Fe_{1.01}Se$, it is clear that the introduction of copper into the structure introduces a local moment. The relative proportions of the doublet and sextet in the Mössbauer data for $Fe_{0.97}Cu_{0.04}Se$ and $Fe_{0.89}Cu_{0.12}Se$ are statistically consistent with the presence of two types of iron site: one in which the iron is neighbour to one or more copper atoms, and at which magnetic fluctuations are observed, and one in which the iron has no neighbouring copper atoms and behaves in an analogous manner to $Fe_{1.01}Se$, with no evidence for long term magnetic fluctuations. As the copper content is increased, the relative proportion of the former iron site, and therefore the local moment, grows until eventually, for $Fe_{0.89}Cu_{0.12}Se$, spin glass-type magnetic ordering is observed. It is not possible to discern from these measurements whether the moment is localized on the copper itself or on the neighbouring iron.

**Conclusions**

Our results show that significant electron doping of tetragonal iron selenide is possible without substantive structural changes, through the substitution of copper at the iron site. Up to 20-30% of the iron can be replaced, before the system moves to adopt a more three-dimensional structure, and this corresponds to a doping of 0.6-0.9 electrons per formula unit – substantially more than has previously been achieved in this system.

It has previously been shown that superconductivity in tetragonal iron selenide is not robust, and sensitive to very slight changes in composition and disorder [8]. Here, we see that it is similarly fragile to electron doping through Fe-site metal substitution, with bulk superconductivity being completely suppressed by as little as 0.045 electrons per mole. The addition of further electrons pushes the system through a metal-insulator transition.

Electron doping has a similarly profound effect on the magnetism of this system. Whereas the parent FeSe displays no sign of magnetic ordering, a moment is clearly evident in both magnetization measurements and Mössbauer spectra for $Fe_{0.97}Cu_{0.04}Se$, with a statistical distribution strongly implying it to be localized at or around the copper sites. The magnitude of this local moment grows with additional electron doping, until longer range, glassy magnetic interactions are finally observed.

Studies on the copper selenides $Cu_xSe$ and $CuSe_2$ [25] have shown the stoichiometric materials to be non-magnetic, so it seems very likely that the magnetic moment in $Fe_{1.01-x}Cu_xSe$ would be localized on the iron. While magnetic order at the iron site is a common feature in studies of the superconducting iron arsenides, this work represents the first observation of such magnetism in the parent compound, $Fe_{1.01}Se$.

**Acknowledgements**

The work at Princeton was supported by the US Department of Energy, Division of Basic Energy Sciences, Grant DE-FG02-98ER45706. T. M. McQueen gratefully acknowledges support of the National Science Foundation Graduate Research Fellowship program.

.

**Figure Captions**

**Figure 1** Variation of high field (9 T) magnetization of $Fe_{1.01-x}Cu_xSe$ and $Cu_xFe_{1.01}Se$ with x, displayed as a proportion of the signal from metallic iron. Inset shows variation of magnetization of $Cu_xFe_{1.01}Se$ with magnetic field.

**Figure 2** (a) Laboratory powder X-ray diffraction patterns for $Fe_{1.01-x}Cu_xSe$. The asterisks mark the largest peak of the major impurity phase, $Fe_7Se_8$. (b) Rietveld refinements of small regions of the XRD data for $Fe_{1.01-x}Cu_xSe$ (x = 0.2, 0.3, 0.5) to illustrate phase separation between FeSe and $CuFeSe_2$ structure types (themselves depicted in the insets)

**Figure 3** (a) Variation of refined lattice parameters with x. (b) Variation of Fe-Se bond distance and upper Se-Fe-Se bond angle with x. Inset illustrates these distances and angles as they make up the $FeSe_4$ tetrahedra.

**Figure 4** Resistivity measurements (normalized to room temperature values) show the transition from superconducting to insulating behaviour as x increases in $Fe_{1.01-x}Cu_xSe$. Inset shows low field susceptibility data for $Fe_{1.01}Se$ and $Fe_{0.995}Cu_{0.015}Se$, illustrating the rapid suppression of superconductivity by only 1.5 % copper substitution.

**Figure 5** Variation of magnetic susceptibility of the $Fe_{1.01-x}Cu_xSe$ samples with temperature, measured between 2 and 4 T in order to negate the effect of ferromagnetic impurities. Upper inset shows behaviour of values extracted at 10 and 298 K with x. Lower inset shows FC and ZFC magnetic susceptibility, measured at 1 T, for $Fe_{0.89}Cu_{0.12}Se$. A clear increase in the local moment with x is observed, culminating in a spin glass-type transition at $Fe_{0.89}Cu_{0.12}Se$, which then broadens out at higher dopings.

**Figure 6** $^{57}Fe$ Mössbauer spectra at 4.2 K. The spectra for $Fe_{0.99}Cu_{0.02}Se$ can be described by a single quadrupole doublet, however the spectra for $Fe_{0.97}Cu_{0.04}Se$ and $Fe_{0.89}Cu_{0.12}Se$ require the inclusion of an additional sextet term.

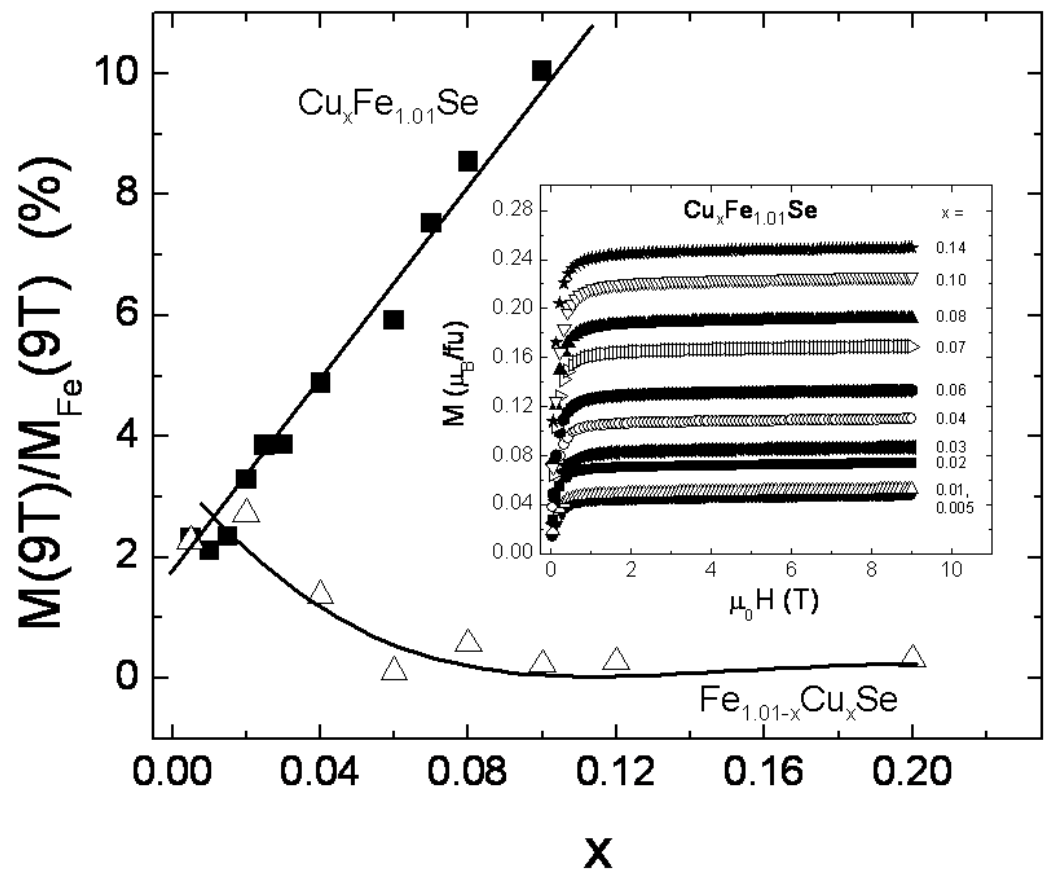

Figure 1

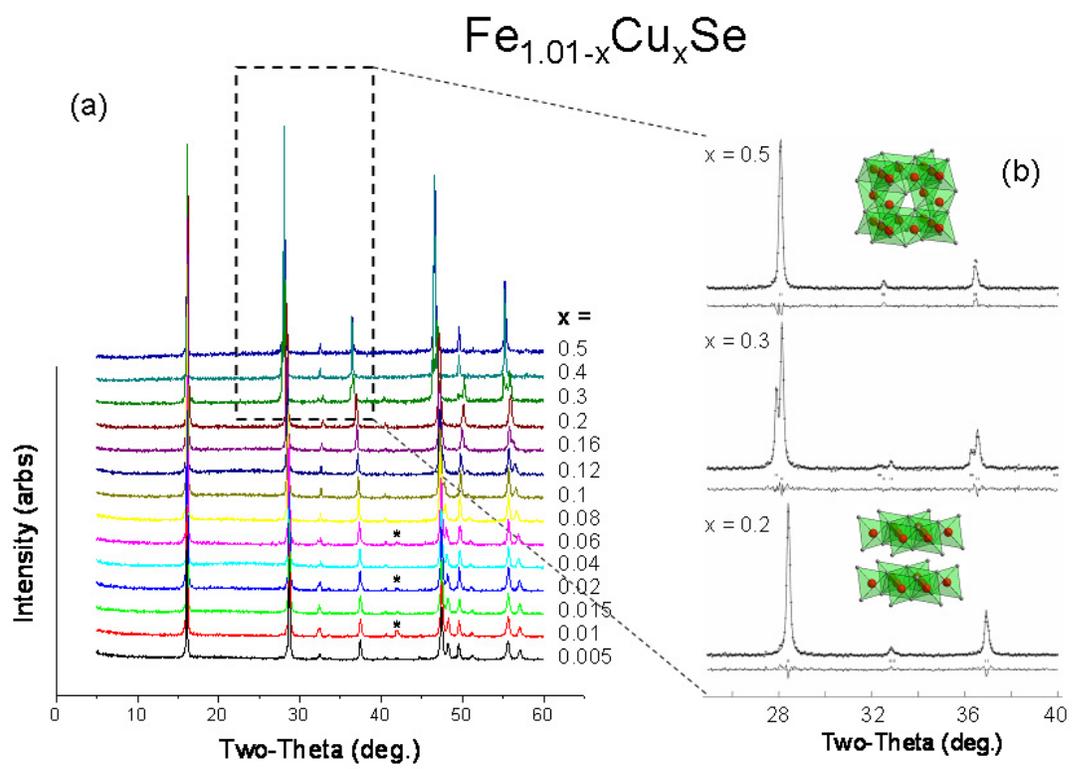

Figure 2

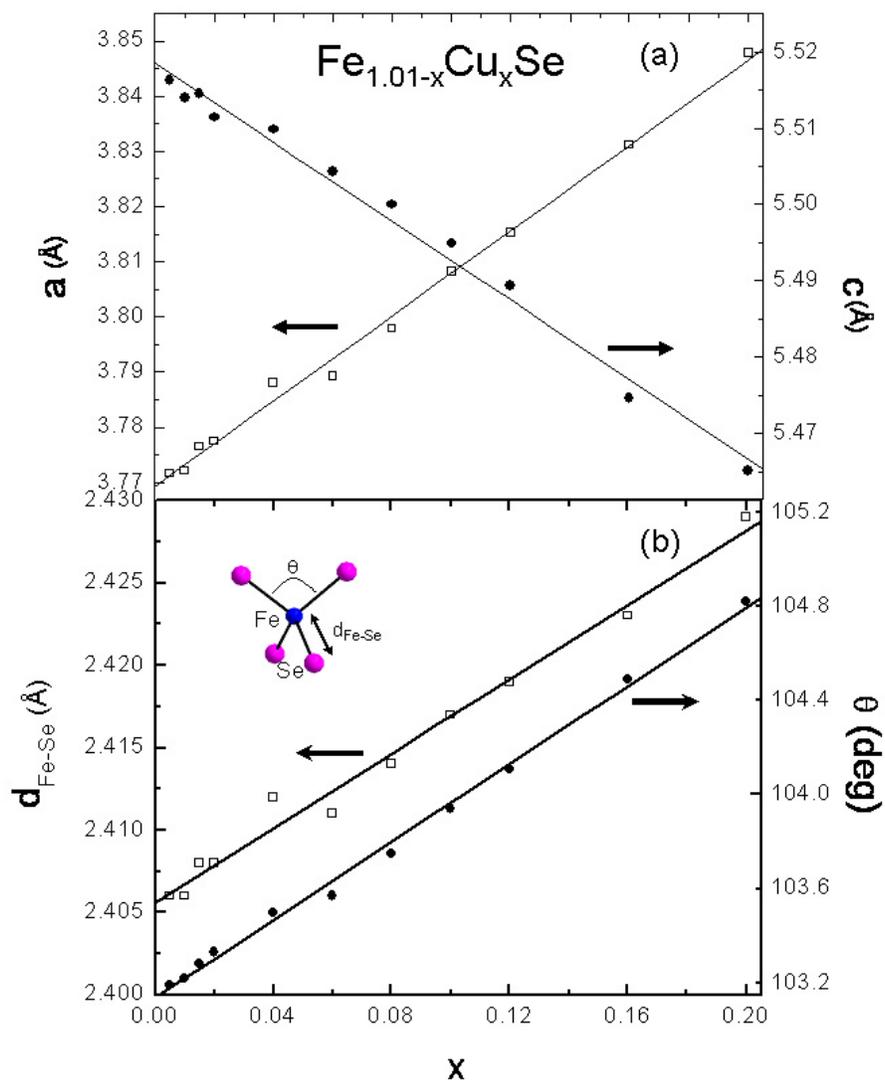

Figure 3

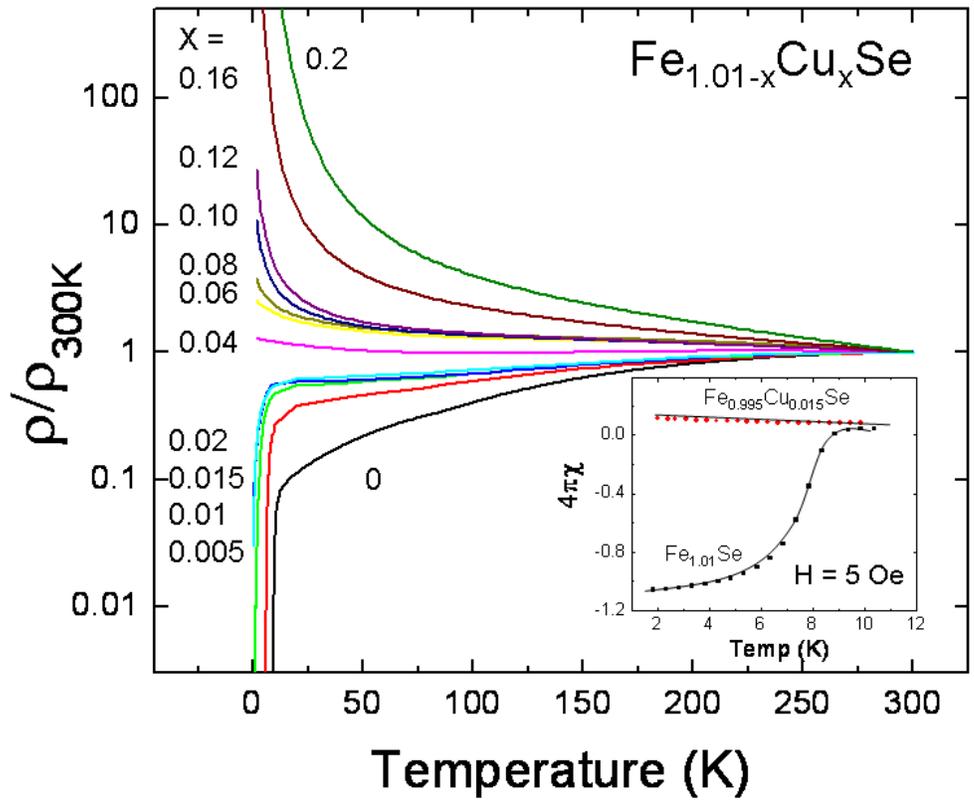

Figure 4

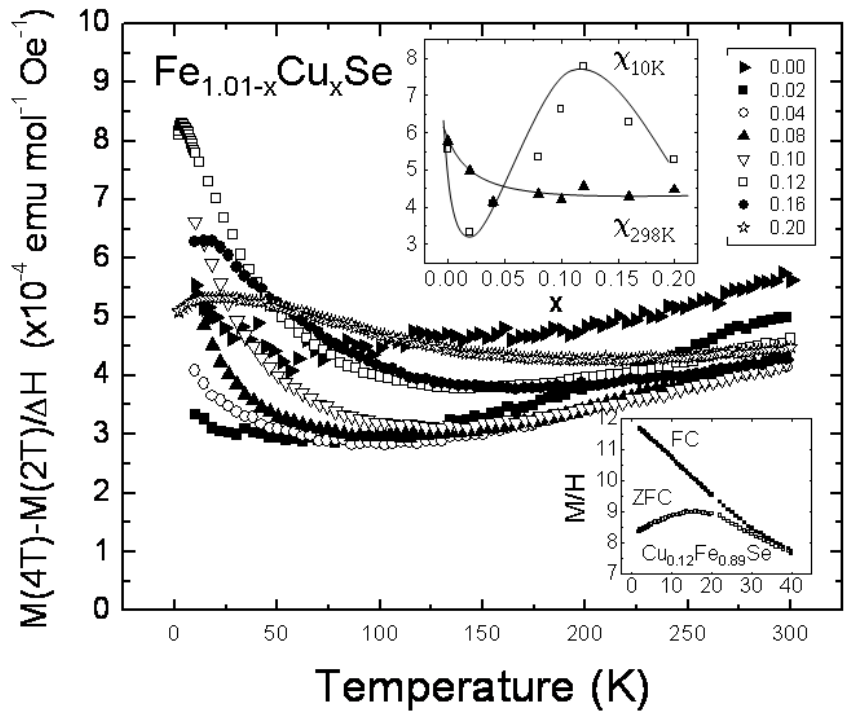

Figure 5

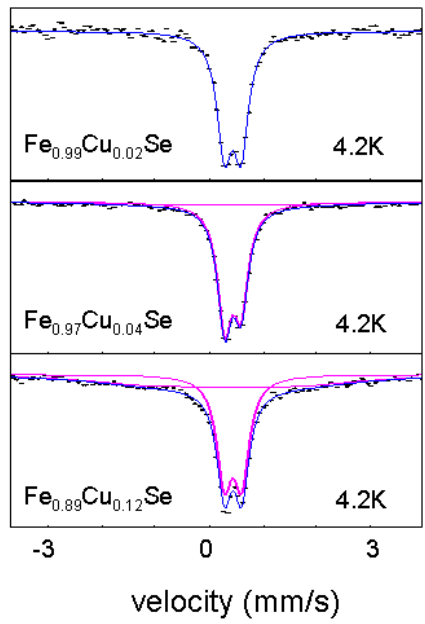

Figure 6